\documentclass[prl,twocolumn,groupedaddress,showpacs]{revtex4}

\usepackage{bm}
\usepackage{graphicx}
\usepackage{color}
\usepackage{ulem}

\begin{document}

\title{Electronic structure of Bi$_2$Te$_3$/FeTe heterostructure: implications for unconventional superconductivity}

\author{Kenta Owada,$^1$ Kosuke Nakayama,$^1$ Ryuji Tsubono,$^1$ Koshin Shigekawa,$^1$ Katsuaki Sugawara,$^{1,2,3}$ Takashi Takahashi,$^{1,2,3}$ and Takafumi Sato$^{1,2,3}$}

\affiliation{$^1$Department of Physics, Tohoku University, Sendai 980-8578, Japan\\
$^2$Center for Spintronics Research Network, Tohoku University, Sendai 980-8577, Japan\\
$^3$WPI Research Center, Advanced Institute for Materials Research, Tohoku University, Sendai 980-8577, Japan
}

\date{\today}

\begin{abstract}
We have performed angle-resolved photoemission spectroscopy on a heterostructure consisting of topological insulator Bi$_2$Te$_3$ and iron chalcogenide FeTe fabricated on SrTiO$_3$ substrate by molecular-beam-epitaxy technique. This system was recently found to show superconductivity albeit non-superconducting nature of each constituent material. Upon interfacing FeTe with two quintuple layers of Bi$_2$Te$_3$, we found that the Dirac-cone surface state of Bi$_2$Te$_3$ is shifted toward higher binding energy, while the holelike band at the Fermi level originating from FeTe moves toward lower binding energy. This suggests that electron charge transfer takes place from FeTe to Bi$_2$Te$_3$ through the interface. The present result points to importance of hole-doped FeTe interface for the occurrence of unconventional superconductivity.
\end{abstract}


\maketitle
Superconductivity associated with an interface of two parent materials is recently attracting tremendous attention because it often becomes a platform of unconventional superconductivity owing to the peculiar characteristic of interface that has circumstances different from bulk. One such example is a charge accumulation at the interface and the resultant emergence of metallicity and superconductivity, as highlighted by the observation of superconductivity at the interface of insulating LaAlO$_3$ (LAO) and SrTiO$_3$ (STO) \cite{LAOSTO1, LAOSTO2}. The interface also plays a crucial role in controlling the superconducting transition temperature, $T_{\rm c}$, as recognized from a significant enhancement of $T_{\rm c}$ in FeSe on STO (over 65 K) compared to that in bulk FeSe (8 K), associated with a charge transfer across the interface and an interfacial electron-phonon coupling \cite{FeSeSTO1, FeSeSTO2, FeSeSTO3}. Also, a widely used strategy to realize topological superconductivity relies on interfacing conventional superconductors and nano-wires/ultrathin films to utilize superconducting proximity effect though the interface. Thus, it is important to find a new platform of interfacial superconductivity and establish the role of interface to unusual superconducting properties.

Recently, it has been reported that a heterostructure consisting of topological insulator Bi$_2$Te$_3$ and a parent compound of iron-chalcogenide superconductors FeTe hosts interfacial superconductivity \cite{BiTeFeTe1}. Whereas both Bi$_2$Te$_3$ and FeTe are non-superconducting, superconductivity appears around 2 K after interfacing one quintuple layer (QL) of Bi$_2$Te$_3$ with a thick FeTe film. The $T_{\rm c}$ increases with increasing the number of Bi$_2$Te$_3$ QLs, and reaches the highest value of $\sim$ 12 K at 5 QL \cite{BiTeFeTe1}, which is about ten times higher than that of a prototypical interfacial superconductor LAO/STO \cite{LAOSTO2}. Also, two-dimensional (2D) character of superconductivity was verified by observations of a Berezinsky-Kosterlitz-Thouless transition and the temperature dependence of upper critical field that follows the Ginzburg-Landau theory for 2D superconductor films \cite{BiTeFeTe1}. These findings triggered intensive investigations on Bi$_2$Te$_3$/FeTe heterostructure \cite{BiTeFeTe2, BiTeFeTe3, BiTeFeTe4, BiTeFeTe5, BiTeFeTe6, BiTeFeTe7} and, consequently, several peculiar properties such as coexistence of multiple gaps in the superconducting state \cite{BiTeFeTe6} and enhancement of $T_{\rm c}$ up to 20 K under hydrostatic pressure \cite{BiTeFeTe7} have been reported. However, the origin of interfacial superconductivity in Bi$_2$Te$_3$/FeTe is still under debate partly because of the lack of detailed understanding of the electronic states. Besides basic interests in the interfacial superconductivity, Bi$_2$Te$_3$/FeTe is also attracting attention as a new topological-superconductor candidate. If the topological Dirac-cone band of Bi$_2$Te$_3$ is preserved at the surface or interface, the heterostructure of Bi$_2$Te$_3$/FeTe would provide a rare opportunity to search for Majorana fermions at relatively high temperatures. To clarify the origin of interfacial superconductivity and the possibility of topological superconductivity, the experimental determination of the electronic states in Bi$_2$Te$_3$/FeTe is of crucial importance.

In this paper, we report angle-resolved photoemission spectroscopy (ARPES) study on Bi$_2$Te$_3$/FeTe heterostructure grown by molecular-beam epitaxy (MBE). We directly observed the evolution of the electronic states upon interfacing Bi$_2$Te$_3$ with FeTe. We found the occurrence of a charge transfer through the interface and the presence of a topological Dirac-cone band derived from Bi$_2$Te$_3$. We discuss the implications of the present results for the origin of interfacial superconductivity and the topological property in Bi$_2$Te$_3$/FeTe.

\begin{figure}
\includegraphics[width=3in]{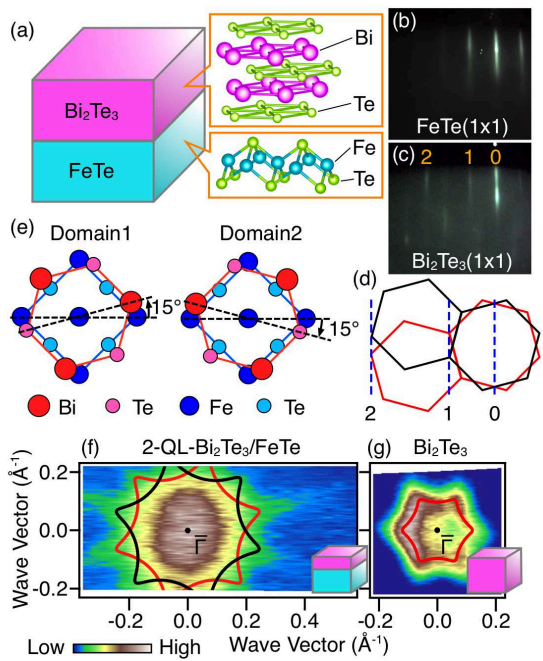}
\vspace{0cm}
\caption{(Color online) (a) Schematic view of $n$-QL Bi$_2$Te$_3$/FeTe heterostructure on STO substrate. (b), (c) RHEED patterns for 10-ML FeTe and 2-QL-Bi$_2$Te$_3$/10-ML-FeTe, respectively. (d) BZ for two types of Bi$_2$Te$_3$ domains (red and black hexagons), estimated from the RHEED pattern in (c). (e) Schematic atomic arrangement of two types of Bi$_2$Te$_3$ crystal domains on FeTe rotated by 30$^\circ$ from each other. (f), (g) ARPES-intensity mapping at $E_{\rm F}$ measured at $T$ = 30 K as a function of 2D wave vector, $k_x$ and $k_y$, for 2-QL-Bi$_2$Te$_3$/10-ML-FeTe/STO and 6-QL-Bi$_2$Te$_3$/Si(111), respectively. Intensity at $E_{\rm F}$ was obtained by integrating the spectra within $\pm$10 meV of $E_{\rm F}$.}
\end{figure}

Heterostructures of Bi$_2$Te$_3$ and FeTe films were fabricated on a Nb (0.05 wt$\%$)-doped STO substrate with the MBE method. The substrate was first degassed at 600 $^\circ$C for 5 h and then heated at 900 $^\circ$C for 30 min. Next, a 10-monolayer (ML) FeTe film was grown by co-evaporating Fe and Te in a Te-rich condition while keeping the substrate temperature at 270 $^\circ$C. Finally, we fabricated $n$-QL Bi$_2$Te$_3$ film ($n$ = 2 and 6) on the FeTe film by co-evaporating Bi and Te at the substrate temperature of 270 $^\circ$C. We also fabricated a 6-QL Bi$_2$Te$_3$ film on Si(111) as a reference. After the growth, films were annealed under ultrahigh vacuum and characterized by $in$-$situ$ ARPES measurements. ARPES measurements were performed using Scienta-Omicron SES2002 and MBS-A1 spectrometers with He and Xe discharge lamps ($h\nu$ = 21.218 and 8.437 eV, respectively) at Tohoku University. The energy and angular resolutions were set to be 7-30 meV and 0.2$^\circ$, respectively.

First, we present characterization of Bi$_2$Te$_3$/FeTe heterostructure. Figure 1(b) shows the reflection high-energy electron diffraction (RHEED) pattern of 10-ML FeTe on a STO(001) substrate. We clearly observe the 1$\times$1 streak pattern originating from FeTe. After co-depositing Bi and Te atoms onto FeTe, the RHEED intensity from FeTe disappears, and a new sharp streak pattern originating from Bi$_2$Te$_3$ appears [Fig. 1(c)]. Besides the (000) reflection marked by ``0", there exist two types of streak patterns, marked by ``1" and ``2". As shown in Fig. 1(d), they are well explained in terms of two types of hexagonal Brillouin zones (BZs) originating from Bi$_2$Te$_3$, rotated by $\pm$15$^\circ$ with respect to the $\Gamma$M line of square BZ in FeTe. As schematically shown in the atomic arrangement in Fig. 1(e), this corresponds to two types of Bi$_2$Te$_3$ crystal domains rotated by 30$^\circ$ from each other, which is naturally expected from the symmetry difference between Bi$_2$Te$_3$ (C$_6$) and FeTe (C$_4$). Such mixture of two domains is also reflected in the Fermi-surface (FS) mapping in Fig. 1(f), which shows a twelve-fold-symmetric intensity distribution. This is well explained in terms of overlap of two types of snowflake-like FSs of Bi$_2$Te$_3$ rotated by 30$^\circ$ from each other [as a reference, the snowflake-like FS of single-domain Bi$_2$Te$_3$ on Si(111) is shown in Fig. 1(g)]. As shown in Fig. 1(f), the FS of Bi$_2$Te$_3$ on FeTe is slightly expanded as compared to that on Si(111). We will come back to this point later.

\begin{figure}
\includegraphics[width=3in]{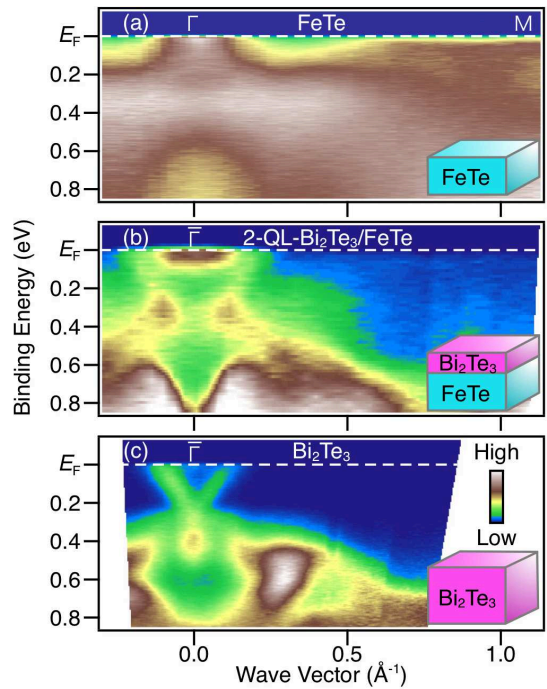}
\vspace{0cm}
\caption{(Color online) (a)-(c) Plots of ARPES intensity in the valence-band region for 10-ML FeTe, 2-QL-Bi$_2$Te$_3$/10-QL-FeTe, and 6-QL-Bi$_2$Te$_3$/Si(111), respectively. Intensities in (a) and (b) were measured along the $\Gamma$M cut of FeTe BZ (15$^\circ$ rotated from the $\bar{\Gamma}\bar{K}$/$\bar{\Gamma}\bar{M}$ cut of Bi$_2$Te$_3$ BZ) with the He-I$\alpha$ line ($h\nu$ = 21.218 eV), whereas that in (c) was measured along the $\bar{\Gamma}\bar{M}$ cut of Bi$_2$Te$_3$ BZ with the Xe-I line ($h\nu$ = 8.437 eV).}
\end{figure}

\begin{figure*}
\includegraphics[width=6.4in]{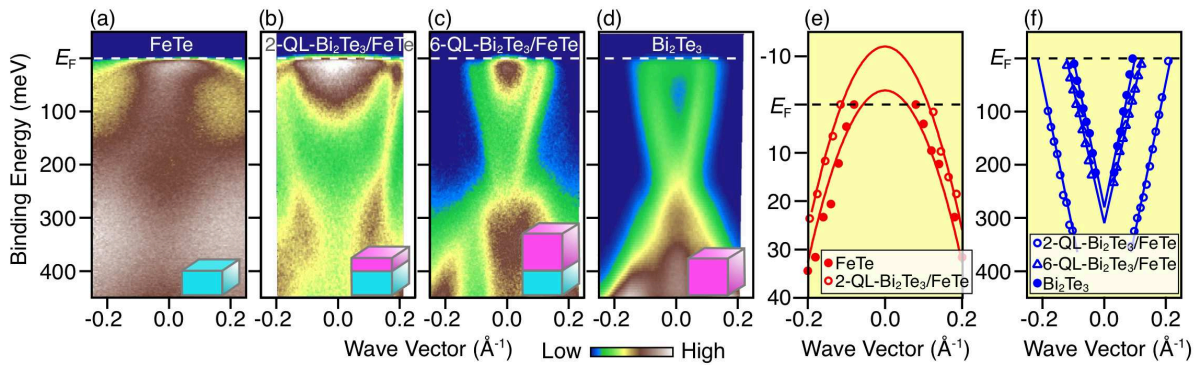}
\vspace{0cm}
\caption{(Color online) (a)-(d) Plots of near-$E_{\rm F}$ ARPES intensity for 10-ML FeTe, 2-QL-Bi$_2$Te$_3$/10-ML-FeTe, 6-QL-Bi$_2$Te$_3$/10-ML-FeTe, and 6-QL-Bi$_2$Te$_3$/Si(111), respectively. (e) Comparison of FeTe-originated holelike band dispersion extracted from the EDCs in (a) and (b). (f) Comparison of Dirac-like band dispersion extracted from the EDCs in (b)-(d).}
\end{figure*}

Having established the orientation of Bi$_2$Te$_3$ on FeTe, next we present the electronic states. Figure 2(a) displays the plot of ARPES intensity at $T$ = 30 K measured along the $\Gamma$M cut of FeTe BZ. One can see three kinds of prominent spectral features; an intense weight in the vicinity of $E_{\rm F}$ at the $\Gamma$ point, a broad feature with a relatively flat dispersion at $\sim$ 0.35 eV around the $\Gamma$ point, and a flat dispersion at the binding energy ($E_{\rm B}$) of $\sim$ 0.05 eV around the M point. This spectral feature is similar to that of bulk FeTe \cite{FeTeARPES1, FeTeARPES2, FeTeARPES3}, confirming its FeTe origin. Upon fabrication of 2-QL Bi$_2$Te$_3$ film on FeTe, we found a drastic change in the spectral feature. Figure 2(b) shows the ARPES intensity for this heterostructure measured along the same ${\bm k}$ cut as Fig. 2(a), which corresponds to the cut rotated by 15$^\circ$ with respect to the $\bar{\Gamma}\bar{K}$/$\bar{\Gamma}\bar{M}$ cut of hexagonal Bi$_2$Te$_3$ BZ. This choice of ${\bm k}$ cut simplifies the data interpretation because the band dispersion from the two crystal domains match each other only along this cut. One can immediately recognize from Fig. 2(b) that the broad feature at $E_{\rm B}$ = 0.3 eV seen in pristine FeTe [Fig. 2(a)] is markedly suppressed. Instead, a new dispersive feature appears at $E_{\rm B}$ = 0.6-0.8 eV around the $\Gamma$ point. Since a similar feature is also seen in pristine Bi$_2$Te$_3$ at similar photon energies \cite{BiTeARPES1, BiTeARPES2}, it is assigned to the Bi$_2$Te$_3$ band. We also find a new holelike band topped at $\sim$ 0.4 eV at the $\Gamma$ point which is also attributed to the Bi$_2$Te$_3$ band, because a similar band appears in Bi$_2$Te$_3$ on Si(111), as displayed in Fig. 2(c).

While the spectral intensity near $E_{\rm F}$ away from the $\Gamma$ point is almost zero in pristine Bi$_2$Te$_3$ [Fig. 2(c)], that in 2-QL-Bi$_2$Te$_3$/FeTe has a finite weight [Fig. 2(b)]. Such feature can be seen in FeTe [Fig. 2(a)], suggesting that a faint photoelectron signal from FeTe beneath 2-QL Bi$_2$Te$_3$ was detected despite significant suppression of the spectral weight. This could be possible because the photoelectron escape depth at this photon energy is $\sim$ 1 nm and as a result about 10 $\%$ of total photoelectrons escape from FeTe through the 2-QL (2 nm)-thick Bi$_2$Te$_3$. We will show later that the observation of buried electronic states of FeTe is corroborated by a close inspection of spectral signature around the $\Gamma$ point near $E_{\rm F}$. The most important spectral signature of 2-QL-Bi$_2$Te$_3$/FeTe in Fig. 2(b) is that there exists a linearly dispersive band across $E_{\rm F}$, reminiscent of the Dirac-cone surface state (SS) in Bi$_2$Te$_3$ in Fig. 2(c). Interestingly, its Fermi vectors appear to be expanded compared to those of pristine Bi$_2$Te$_3$. This change is also responsible for the observed expansion of FS in Figs. 1(f) and 1(g). The experimental fact that the twelve-fold symmetric intensity pattern in Fig. 1(f) originates from the linearly dispersive band in Fig. 2(b) supports that this band is of Dirac-cone SS origin.

To see more clearly the change in the electronic states upon interfacing Bi$_2$Te$_3$ and FeTe, we show in Fig. 3 the near-$E_{\rm F}$ ARPES intensity around the $\Gamma$ point for (a) pristine FeTe (10 ML), (b) 2 and (c) 6 QLs of Bi$_2$Te$_3$ on FeTe, and (d) Bi$_2$Te$_3$ on Si(111), measured with higher statistics and energy resolution. One can see from a side-by-side comparison of Figs. 3(a) and 3(b) that a shallow holelike band that crosses $E_{\rm F}$ around the $\Gamma$ point in FeTe [Fig. 3(a)] is still seen even after growth of 2-QL Bi$_2$Te$_3$ on FeTe [Fig. 3(b)], while its intensity is markedly suppressed. On increasing the number of QLs to 6, the holelike band completely disappears [Fig. 3(c)]. This is reasonable since the photoelectrons cannot escape from buried FeTe because Bi$_2$Te$_3$ is too thick ($\sim$ 6 nm). Another important aspect of Fig. 3 is the Dirac-cone SS. One can see that 6-QL Bi$_2$Te$_3$ on FeTe [Fig. 3(c)] and pristine Bi$_2$Te$_3$ [Fig. 3(d)] shows a similar Dirac-cone feature at similar $E_{\rm B}$'s (note that the conduction band is not clearly seen in pristine Bi$_2$Te$_3$ because the $k_z$ value is not rightly at the conduction-band minimum), whereas that for 2-QL Bi$_2$Te$_3$ on FeTe sinks well below $E_{\rm F}$ [Fig. 3(b)].

\begin{figure}
\includegraphics[width=3in]{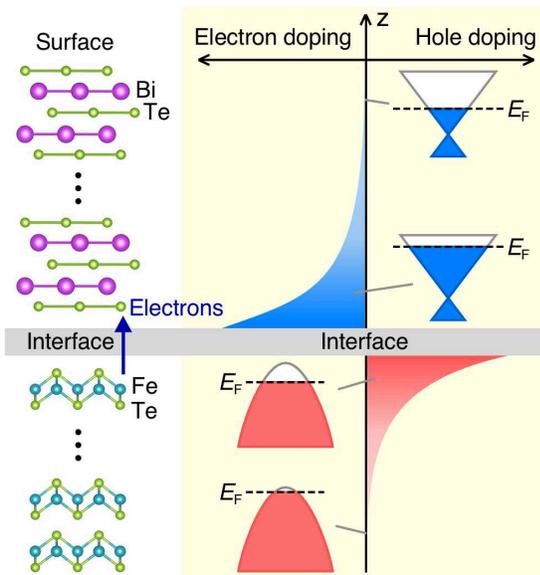}
\vspace{0cm}
\caption{(Color online) Schematic crystal structure, band diagram, and density of extra electron/hole carriers of Bi$_2$Te$_3$/FeTe heterostructure.}
\end{figure}

To discuss quantitatively the observed change in the energy bands, we have estimated the energy position of FeTe-originated holelike band by tracing the peak position of energy distribution curves (EDCs), and the result is shown in Fig. 3(e). It is obvious that the shape of band dispersion is quite similar between 2-QL-Bi$_2$Te$_3$/FeTe and pristine FeTe, whereas the former band is shifted as a whole by 10 meV toward lower $E_{\rm B}$ with respect to the latter one. This indicates that the FeTe surface (interface) is hole-doped upon interfacing the Bi$_2$Te$_3$ layer. By contrast, the downward shift of the Dirac-cone state in 2-QL-Bi$_2$Te$_3$/FeTe compared to pristine Bi$_2$Te$_3$, as clarified in the quantitative analysis of band energies in Fig. 3(f), signifies that the Bi$_2$Te$_3$ layers interfaced with FeTe are electron-doped. A similar energy position of the Dirac cones between 6-QL-Bi$_2$Te$_3$/FeTe and pristine Bi$_2$Te$_3$ shown in Fig. 3(f) implies that extra electron charge is accumulated near the interface. These observations are summarized in the schematic depth profile of band diagram and extra electron/hole charge carriers in Fig. 4. The electron- vs hole-doped nature of Bi$_2$Te$_3$ and FeTe suggests that electron charge transfer from FeTe to Bi$_2$Te$_3$ takes place. This is reasonable since the work function of Bi$_2$Te$_3$ (5.3 eV) \cite{BiTeWF} is smaller than that of FeTe (4.4 eV) \cite{FeTeWF}, so that electrons are transferred from FeTe to Bi$_2$Te$_3$ upon making junction.

Now we discuss implications of the present ARPES results in relation to the occurrence of superconductivity. What is fascinating in this heterostructure is that the superconductivity emerges upon junction of non-superconducting Bi$_2$Te$_3$ and FeTe \cite{BiTeFeTe1}. There may be two possible explanations for the electronic states responsible for this superconductivity; (i) the Bi$_2$Te$_3$/FeTe interface itself hosts a new band structure different from that of parent materials wherein the superconductivity emerges, or (ii) either carrier-doped Bi$_2$Te$_3$ or FeTe itself becomes superconducting around the interface. The present ARPES result strongly suggests that (i) is unlikely, since all the bands observed in 2-QL-Bi$_2$Te$_3$/FeTe [Figs. 2(b) and 3(b)] can be assigned either to the Bi$_2$Te$_3$- or FeTe-originated bands. Then, the next question is which, electron-doped Bi$_2$Te$_3$ or hole-doped FeTe, or both is superconducting. It is empirically known from previous transport and spectroscopic studies of bulk Bi$_2$Te$_3$ that superconductivity does not emerge by electron-doping by replacement and/or intercalation of atoms. This situation is different from another prototypical topological insulator, Bi$_2$Se$_3$, where intercalation of various atoms (Cu, Nb, and Sr) causes electron-doping and triggers superconductivity \cite{CuBiSe, SrBiSe, NbBiSe}. This naturally leads to a conclusion that electron-doped Bi$_2$Te$_3$ does not host superconductivity. Our result thus implies that the hole-doped FeTe interface is responsible for the superconductivity.

Above consideration naturally favors the scenario that topological superconductivity may take place in Bi$_2$Te$_3$. Assuming that the FeTe interface is superconducting, it would be possible to induce a pairing gap at the Dirac-cone surface/interface state in the QLs of Bi$_2$Te$_3$ by the superconducting proximity effect from FeTe, if the Bi$_2$Te$_3$ film is thick enough to realize the Dirac-cone SSs at the top and bottom surfaces. Since the Dirac-cone band was observed even in the 2-QL-thick Bi$_2$Te$_3$ film \cite{BiTeARPES2}, we think that the 2 (and also 6) QL films in the present study can host the spin-helical Dirac-cone states. This would satisfy the theoretically predicted condition of topological superconductivity that utilizes the helical Dirac fermions \cite{TSC}. A next important challenge is to directly determine the pairing symmetry by measuring the ${\bm k}$ dependence of the proximity-induced gap, and search for a spectroscopic signature of Majorana bound state in the vortex core.

In conclusion, we reported ARPES study on a heterostructure of Bi$_2$Te$_3$ and FeTe epitaxially grown on STO. By comparing the band structure among Bi$_2$Te$_3$/FeTe, FeTe, and Bi$_2$Te$_3$, we found that the electron charge transfer from FeTe to Bi$_2$Te$_3$ takes place upon interfacing Bi$_2$Te$_3$ with FeTe. Moreover, the influence of charge transfer was found to be more prominent for a thinner Bi$_2$Te$_3$ film, suggesting the charge accumulation near the interface. Taking into account that electron-doped Bi$_2$Te$_3$ is unlikely to be a superconductor, we suggested that hole-doped FeTe at the interface is responsible for the occurrence of superconductivity in the heterostructure. This points to possible topological superconductivity occurring at the Dirac-cone surface/interface states in Bi$_2$Te$_3$.

\begin{acknowledgments}
We thank Masato Kuno and Takumi Sato for their assistance in the ARPES experiments. This work was supported by JST-PRESTO (No: JPMJPR18L7), JST-CREST (No: JPMJCR18T1), MEXT of Japan (Innovative Area``Topological Materials Science" JP15H05853), JSPS (JSPS KAKENHI No: JP17H04847, JP17H01139, and JP18H01160), and KEK-PF (Proposal number: 2018S2-001).
\end{acknowledgments}

\bibliographystyle{prsty}

\end{document}